\documentstyle[12pt,psfig]{article}
\newcommand{\be}{\begin{equation}}
\newcommand{\ee}{\end{equation}}
\newcommand{\bea}{\begin{eqnarray}}
\newcommand{\eea}{\end{eqnarray}}
\def\aprle{\buildrel < \over {_{\sim}}}
\def\aprge{\buildrel > \over {_{\sim}}}
\begin{document}
\topmargin 0pt
\renewcommand{\thefootnote}{\fnsymbol{footnote}}     
\setcounter{page}{0}
\begin{titlepage}     
\begin{flushright}
IC/97/49\\
hep-ph/9705451
\end{flushright}
\vspace{1.5cm}
\begin{center}
{\large THE NEUTRINO MAGNETIC MOMENT AND 
TIME VARIATIONS OF THE SOLAR NEUTRINO FLUX
\footnote{Invited talk given at the 4th International Solar Neutrino 
Conference, Heidelberg, Germany, April 8--11, 1997}} \\
\vspace{0.4cm}
\vspace{0.5cm}
{\large E.Kh. Akhmedov
\footnote{On leave from RRC ``Kurchatov Institute'', Moscow 123182, Russia.
e-mail address: akhmedov@sissa.it}}\\
\vspace{0.2cm}
{\em International Centre for Theoretical Physics \\
 Strada Costiera 11, I-34100 Trieste, Italy}\\
\vspace{0.5cm}
\end{center}
\vspace{1.4cm} 
\begin{abstract}
The present status of the neutrino magnetic moment solutions of the solar 
neutrino problem is summarized. After a brief review of the basics of the 
neutrino spin and spin--flavor precession I discuss the experimental data 
and show how the neutrino resonant spin--flavor precession (RSFP) mechanism 
can naturally account for sizeable time variations in the Homestake signal 
and no observable time variations in the Kamiokande and gallium experiments. 
Fits of the existing data and predictions for the forthcoming solar neutrino 
experiments are also discussed. In the last section I summarize the objections 
to the RSFP mechanism that are frequently put forward and comment on them. 
\end{abstract}
\end{titlepage}
\vspace{.5cm}
\renewcommand{\thefootnote}{\arabic{footnote}}
\setcounter{footnote}{0}
\newpage

\section{Introduction}
The combined data of all the currently operating solar neutrino 
experiments strongly suggest that the observed deficiency of the 
solar neutrino flux is due to some unconventional neutrino 
properties and not to astrophysics. The analysis \cite{BPBC} also shows 
that the standard solar models which take into account the element 
diffusion are in a very impressive agreement with the high-accuracy 
helioseismological data whereas the so-called unconventional solar 
models grossly disagree with them. This lends further support to 
the point of view that it is some new neutrino physics and not our 
poor knowledge of the solar physics that is responsible for the solar 
neutrino problem.

The main neutrino--physics explanations of the problem that have been 
discussed so far are neutrino oscillations (in solar matter or in vacuum), 
neutrino decay, and neutrino magnetic moments. The relevant process that is 
caused by the neutrino magnetic moment is the neutrino spin precession in the 
solar magnetic field. By now, the neutrino decay explanation has been ruled 
out: It predicts stronger depletion of the lower-energy neutrino flux, in 
direct contradiction with the data. Vacuum and 
matter--enhanced oscillations (the MSW effect) are in a good shape 
\cite{oscill}; the latter is at the moment the most popular candidate for the 
solution of the solar neutrino problem. It is generally believed that the 
neutrino magnetic moment is strongly disfavored by the present data, if not 
excluded. I will try to convince you that this is wrong. It may well be that 
the magnetic moment solutions soon will be ruled out by the Super--Kamiokande 
and SNO experiments; however, at the moment they are in a perfectly good 
agreement with all the existing data. 

The neutrino magnetic moment can account for both the observed deficiency of 
the solar neutrino flux and the time variation of the signal. The time 
dependence may be caused by time variations of the magnetic field in 
the convective zone of the sun. Some evidence for such time variations exists 
in the Homestake data, whereas no time variations were observed in Kamiokande 
and gallium experiments, at least within their experimental errors. I will 
explain that this may be a very natural consequence of the neutrino 
transition (flavor-off-diagonal) magnetic moment. 

\section{Neutrino spin precession}
The idea that neutrino magnetic moment may have something to do with the 
deficiency of the solar neutrino flux was first put forward by Cisneros 
more than 25 years ago \cite{Cis}. His point was very simple: If the 
electron neutrino has a magnetic moment, its spin will precess in a 
transverse magnetic field which means that a fraction of the left-handed 
neutrinos $\nu_{eL}$ will be converted into the right-handed $\nu_{eR}$. 
These $\nu_{eR}$ are sterile -- they do not take part in the usual weak 
interactions and therefore cannot be detected, which explains the deficiency 
of the observed flux. It is not difficult to find the probability of the 
$\nu_{eL}\to\nu_{eR}$ transition: 
\be
P(\nu_{eL}\to\nu_{eR};r)=\sin^2 \left(\int_0^r\mu B_\bot(r')\,dr'\right)\,.
\label{P1}
\ee
Here $\mu$ is the neutrino magnetic moment and $B_\bot(r)$ is the transverse 
magnetic field strength, $r$ being the coordinate along the neutrino path. 
This formula is valid for an arbitrary magnetic field profile $B_\bot(r)$. 
There are two important things to be noticed. First, the transition 
probability does not depend on the neutrino energy, i.e. neutrinos of all 
the energies experience the same degree of the $\nu_{eL}\to\nu_{eR}$ 
conversion, and the spectrum of the surviving $\nu_{eL}$'s is undistorted. 
Second, the amplitude of the precession is unity: If the neutrino magnetic 
moment is large enough, the transverse magnetic field is strong and extended 
enough, in principle a beam of the $\nu_{eL}$'s can be {\em fully} converted 
into the $\nu_{eR}$ beam. 

In his paper \cite{Cis} Cisneros assumed that there is a strong 
constant in time magnetic field in the core of the sun, and this field 
causes the neutrino spin precession. However, he did not take 
into account the matter effects on the precession, which can be quite 
strong in the dense solar core. At that time it was simply not known that 
matter can influence neutrino propagation significantly. The solar matter 
effects on the neutrino spin precession were considered by Voloshin, 
Vysotsky and Okun (VVO) \cite{VVO} and by Barbieri and Fiorentini \cite{BF}. 
VVO also made the following important observation: There is a strong toroidal 
magnetic field in the convective zone of the sun, and this field varies in 
time with the 11-year periodicity. If one assumes that at the maxima of 
solar activity, when one observes the maximum number of sunspots, the 
magnetic field is strongest, one can expect to see the smallest flux of 
solar neutrinos at these periods. This is because the neutrino spin 
precession converting the $\nu_{eL}$'s into unobservable $\nu_{eR}$'s 
should be most efficient for strongest magnetic fields. In other words, the 
observed solar neutrino flux should vary in time in {\em anticorrelation} 
with solar activity. Another interesting point made by VVO was the following: 
It is known that the toroidal magnetic field has opposite directions in the 
northern and southern hemispheres of the sun. This means that it must vanish 
near the solar equator. The orbit of the earth is inclined by  $7^\circ 15'$ 
to ecliptic, therefore twice a year (in the beginning of June and in the 
beginning of December) 
the solar core is seen from the earth through the equatorial gap in the 
toroidal magnetic field of the sun. One can expect the solar neutrinos 
observed on the earth during these periods to be unaffected by the solar 
magnetic filed and therefore their flux not to be suppressed even in the 
periods of high solar activity. This should lead to very peculiar semiannual 
variations of the solar neutrino signal.  I will discuss the experimental 
status of the time variations of the solar neutrino flux later on. 

In the presence of matter one cannot find the precession probability for 
arbitrary magnetic field and matter density profiles in a closed 
analytical form. However, in the simplest case of the uniform magnetic 
field and constant matter density such an expression can be readily found 
\cite{VVO,BF}: 
\be
P(\nu_{eL}\to\nu_{eR};r)=\frac{(2\mu B_\bot)^2}{[\sqrt{2}G_F(N_e-N_n/2)]^2
+(2\mu B_\bot)^2]}\sin^2 
\left(\frac{1}{2}\sqrt{D_1}\, r \right)\,. 
\label{P2}
\ee
Here $G_F$ is the Fermi constant, $N_e$ and $N_n$ are the electron 
and neutron number densities, and $D_1$ in the argument of the sine 
is the denominator of the pre-sine factor. Comparing this expression 
with eq. (\ref{P1}), we see an important difference: in matter the 
precession amplitude (i.e. the pre-sine factor) is always less than one. 
The reason for that is very simple. In vacuum the left-handed and 
right-handed components of a Dirac neutrino are degenerate, so the precession 
relates the states of the same energy and therefore proceeds with the 
amplitude equal to unity. In matter this is no longer true -- left-handed 
neutrinos experience coherent forward scattering on the particles of the
medium, i.e. receive some mean potential energy, whereas the sterile 
right-handed neutrinos have zero potential energy. This results in an 
energy splitting of the $\nu_{eL}$ and $\nu_{eR}$. The neutrino spin 
precession now relates the states of different energy and its amplitude is 
therefore always less than one. If the interaction that mixes the $\nu_{eL}$ 
and $\nu_{eR}$, the Zeeman energy $\mu B_\bot$, is much smaller than the 
energy splitting $\sqrt{2}G_F(N_e-N_n/2)$, the precession is strongly 
suppressed. 

There is one thing that is common for eqs. (\ref{P1}) and (\ref{P2}): in 
both cases the transition probability does not depend on the neutrino 
energy, i.e. neutrinos of all the energies experience the same degree 
of the $\nu_{eL}\to \nu_{eR}$ conversion. This is an important property of 
the neutrino spin precession that holds for arbitrary magnetic field and 
matter density profiles. 

\section{Resonant spin-flavor precession of neutrinos}
\subsection{General features}
If the lepton flavor is not conserved, neutrinos may have flavor-off-diagonal 
(transition) magnetic moments. Such magnetic moments would cause, e.g., 
radiative neutrino decays $\nu_2 \to\nu_1+\gamma$. In transverse magnetic 
fields neutrinos with transition magnetic moments will experience a very 
special type of the spin precession: their spin will rotate with their 
favor changing simultaneously \cite{SchV,VVO}. This is called the 
spin--flavor precession (SFP). For example, if neutrinos are the Dirac 
particles, $\nu_{eL}$ can be rotated into sterile $\nu_{\mu R}$. Majorana 
neutrinos cannot have ordinary magnetic moments because of CPT invariance; 
however, they can have the transition magnetic moments. For Majorana 
neutrinos the SFP converts left-handed neutrinos of a given flavor into 
right-handed {\em antineutrinos} of a different flavor. The latter are 
not sterile -- they are just the usual antineutrinos which take part   
in the standard weak interactions. It is not difficult to find the 
probability of the SFP in vacuum. Consider, for example, the $\nu_{eL}
\leftrightarrow \bar{\nu}_{\mu R}$ transitions. In the case of the uniform 
magnetic field the transition probability for relativistic neutrinos is  
\be
P(\nu_{eL}\to\bar{\nu}_{\mu R};r)=\frac{(2\mu B_\bot)^2}
{(\Delta m^2/2E)^2 +(2\mu B_\bot)^2}\sin^2 \left(\frac{1}{2}\sqrt{D_2}\, 
r \right)\,. 
\label{P3}
\ee
Here $\mu$ is now the transition magnetic moment, $\mu\equiv \mu_{\nu_e 
\nu_\mu}$, $\Delta m^2=m_{\nu_\mu}^2-m_{\nu_e}^2$, and $D_2$ is the 
denominator of the pre-sine factor. For simplicity I have assumed here that 
the $\nu_{e}$ and $\nu_{\mu}$ are mass eigenstates; I will discuss the 
case when both transition magnetic moment and neutrino mass mixing are 
present later. 

For some time the SFP of neutrinos did not attract much attention. The 
reason for this was that its probability is suppressed even in vacuum. 
Indeed, neutrinos of different flavor are expected to have different masses, 
so even in the absence of matter we now have the transitions between the 
non-degenerate neutrino states. The kinetic energy difference of the 
relativistic neutrinos is $\Delta m^2/2E$; if this energy difference is 
small compared to the ``Zeeman energy'' $\mu B_\bot$ the precession 
amplitude is suppressed. However, in 1988 it was realized that in matter 
the situation changes drastically \cite{LM,Akhm1}. Left-handed neutrinos 
of a given flavor and right-handed neutrinos or antineutrinos of a 
different flavor experience different coherent forward scattering on the 
particles of the medium, and so there is a potential energy difference 
which has to be subtracted from the kinetic energy difference in the 
expression for the transition probability:
\be
P(\nu_{eL}\to\bar{\nu}_{\mu R};r)=\frac{(2\mu 
B_\bot)^2}{(\Delta m^2/2E-\sqrt{2}G_FN_{\rm eff})^2 +(2\mu B_\bot)^2}
\sin^2 \left(\frac{1}{2}\sqrt{D_3}\, r \right)\,. 
\label{P4}
\ee
Here $D_3$ is the denominator of the pre-sine factor and $N_{\rm eff}$
depends on the nature of the neutrinos: 
\bea
N_{\rm eff}&=& N_e-N_n/2 \;\; {\mbox {\rm for Dirac neutrinos}}\nonumber \\
           &=& N_e-N_n  \;\; \;\;\;  {\mbox {\rm for Majorana neutrinos}}
\label{Neff}
\eea
Now, for a given $\Delta m^2$ and any value of the neutrino energy $E$ there 
is a certain value of $N_{\rm eff}$ for which the potential energy difference 
$\sqrt{2}G_F N_{\rm eff}$ exactly cancels the kinetic energy difference 
$\Delta m^2/2E$, and the amplitude of the SFP becomes equal to unity, no 
matter how small the transition magnetic moment $\mu$ and how weak the 
magnetic field $B_\bot$. The precession amplitude as a function of 
$N_{\rm eff}$ has a typical resonant behavior, i.e. matter can resonantly 
enhance the neutrino SFP. Of course, having a large precession amplitude is 
not sufficient for the precession probability to be large; the phase of 
the sine in eq. (\ref{P4}) should also be not too small. This puts a 
lower limit on the product $\mu B_\bot$. In the matter with varying 
density the lower bound on $\mu B_\bot$ is imposed by the adiabaticity 
condition which I will shortly discuss. 

The resonant spin--flavor precession (RSFP) of neutrinos in matter is very 
similar to the resonant enhancement of the neutrino oscillations in matter -- 
the celebrated MSW effect \cite{MS,W}. In fact, the mathematics describing 
the neutrino oscillations and spin (or spin--flavor) precession is just the 
same (although there are some important differences in the physics involved). 
This is not by accident, of course. The processes of both types in the 
simplest case of two neutrino species involved reduce to a well-known quantum 
mechanical problem of a two-level system in an external field; one can find a 
discussion of this problem in any textbook of quantum mechanics. The 
evolution of the system is governed by the Shroedinger-like equation
\be
i\frac{d}{dr}\left(\begin{array}{l}
   \nu_{i}\\
   \nu_{j}
\end{array}\right )
~=~\left (
\begin{array}{cc}
   E_i  & V_{ij} \\  
   V_{ij}^* & E_j
\end{array}\right )\left(\begin{array}{l}
   \nu_{i}\\
   \nu_{j}\\
\end{array}\right)
\label{evol1}
\ee
Here $\nu_i$ and $\nu_j$ are the probability amplitudes of finding the 
corresponding neutrino at a point $r$, $E_i$ and $E_j$ are the energies   
of the neutrinos in the absence of mixing, and $V_{ij}$ is the mixing 
interaction. The weak eigenstate neutrinos $\nu_i$ and $\nu_j$ can be 
of the same or different flavor and of the same of different chirality, 
depending on the process in question. For the neutrino spin precession they  
are of the same flavor but different chirality, for the neutrino oscillations 
they are of different flavor and the same chirality, and for the SFP they are 
of different flavor and different chirality. 
The mixing interaction $V_{ij}$ also depends on the process in question.
For the neutrino spin or spin--flavor precession it is $\mu B_\bot$ where 
$\mu$ is the ordinary or transition magnetic moment respectively; for the 
neutrino oscillations $V_{ij}=(\Delta m_{ij}^2/4E)\sin 2\theta_0$ where 
$\theta_0$ is the mass mixing angle (i.e. the angle that relates mass 
eigenstates with flavor eigenstates). For relativistic neutrinos, the 
diagonal matrix elements of the effective Hamiltonian in eq. (\ref{evol1}) 
are $E_i\simeq E+m_i^2/2E+V(\nu_i)$ where $V(\nu_i)$ is the matter-induced 
mean potential energy of the neutrino 
$\nu_i$ \cite{W}:
\be
V(\nu_{eL})=-V({\bar\nu}_{eR})=\sqrt{2}G_F(N_e-N_n/2)\,,
\label{Ve}
\ee
\be
V(\nu_{\mu L},\nu_{\tau L})=-V(\bar{\nu}_{\mu R},\bar{\nu}_{\tau R})=
\sqrt{2}G_F(-N_n/2)\,.
\label{Vmu}
\ee
For sterile right-handed neutrinos and their left-handed antiparticles 
$V(\nu_i)=0$. 

The transition probabilities depend on the difference of the phases 
of the neutrino states. This means that it is only the difference of the 
diagonal elements of the effective Hamiltonian in eq. (\ref{evol1}) (and not 
the diagonal elements themselves) that matters: 
\be
\Delta E_{ij} = \Delta m_{ij}^2/2E +V(\nu_i)-V(\nu_j)\,.
\label{Eij}
\ee
Thus, the neutrino energy $E$ enters into the evolution equation only through 
the parameter $\Delta m_{ij}^2/2E$. For the neutrinos of the same flavor 
$\Delta m_{ij}^2=0$; this explains why the transition probability for the 
ordinary (flavor--conserving) neutrino spin precession is independent of 
the neutrino energy. The effective potential energy term $\sqrt{2}G_F 
N_{\rm eff}$ in eq. (\ref{P4}) is nothing but the difference 
$V(\nu_i)-V(\nu_j)$ [compare eqs. (\ref{Neff}), (\ref{Ve}) and (\ref{Vmu})]. 

The transition probabilities for the neutrino spin and spin--flavor 
precession in the absence of matter or in a matter of constant density, 
(\ref{P1})--(\ref{P4}), can be easily obtained from the evolution equation 
(\ref{evol1}). However in general, in a medium with arbitrary matter 
density and magnetic field profiles, no analytical closed--form expression 
for the transition probability can be obtained. In this case one has to 
solve eq. (\ref{evol1}) numerically, which is quite straightforward. However, 
even without performing any numerical calculations one can get an important 
insight into the evolution of the neutrino system in the adiabatic 
approximation. Consider, for example, the SFP -- induced transitions 
$\nu_{eL}\leftrightarrow \bar{\nu}_{\mu R}$. At any point $r$ one can 
diagonalize the effective Hamiltonian in eq. (\ref{evol1}) and find the 
instantaneous eigenstates 
\bea
\nu_A&=& \cos\theta \;\nu_{eL}+\sin\theta \;\bar{\nu}_{\mu R}\nonumber \\
\nu_B&=& -\sin\theta \;\nu_{eL}+\cos\theta \;\bar{\nu}_{\mu R}
\label{eigen}
\eea
where the mixing angle $\theta$ is defined through  
\be
\tan 2\theta =\frac{2\mu B_\bot}{\sqrt{2}G_F(N_e-N_n)-\Delta m^2/2E}\,.
\label{tan}
\ee
Thus the neutrino eigenstates in matter and magnetic fileds are the linear 
combinations of the neutrinos of different flavor and chirality. 
It should be clearly understood that the mixing angle $\theta$ in 
eqs. (\ref{eigen}) and (\ref{tan}) has nothing to do with the usual mixing 
angle $\theta_0$ that governs the neutrino oscillations. For the moment I 
assume $\theta_0=0$. 

The resonance condition for the SFP is 
\be
\sqrt{2}G_F(N_e-N_n)=\Delta m^2/2E\,.
\label{res1}
\ee
At the resonance point the mixing angle $\theta=\pi/4$ and the amplitude 
of the SFP in eq. (\ref{P4}) becomes equal to unity 
(remember that $N_{\rm eff}=N_e-N_n$ for Majorana neutrinos). 
The resonance condition (\ref{res1}) is very similar to that 
for the MSW effect \cite{MS,W} 
\be
\sqrt{2}G_F N_e =(\Delta m^2/2E)\cos 2\theta_0 \,.
\label{res2}
\ee
The main difference is the $-N_n$ term which is due to the neutral current 
interactions of neutrinos with matter. It is absent in eq. (\ref{res2}) but 
enters into eq. (\ref{res1}) because the SFP, unlike the neutrino 
oscillations, is sensitive to the neutral currents. 
The resonance of the SFP takes place in the $\nu_{eL}\leftrightarrow 
\bar{\nu}_{\mu R}$ channel if $\Delta m^2\equiv m_{\nu_\mu}^2-m_{\nu_e}^2>0$ 
and in the $\bar{\nu}_{eR}\leftrightarrow \nu_{\mu L}$ 
channel if $\Delta m^2<0$. I will be assuming that $\Delta m^2>0$. 

Consider the variation of the mixing angle $\theta$ along the path of 
the solar neutrinos. In the core of the sun where they are born the 
density is high. Assume that it is much larger than the resonance density 
defined through eq. (\ref{res1}). Then the denominator of eq. (\ref{tan}) is 
large, i.e. the mixing angle is small: $\theta(r\simeq 0)\simeq 0$.
With decreasing density the mixing angle $\theta$ increases; it reaches 
$\pi/4$ at the  resonance point and continues to increase. 
Close to the surface of the sun the matter density is very small and one can 
neglect it as compared to the $\Delta m^2/2E$ term in the denominator of 
eq. (\ref{tan}). At the same time, the magnetic field strength decreases 
towards the surface of the sun, so at $r=R_\odot$ the tangent of 
$2\theta$ is a small negative number, i.e. $\theta\simeq \pi/2$. Thus, as 
neutrino propagates from the core of the sun to its surface, $\theta$ changes 
from $\simeq 0$ to $\simeq \pi/2$, passing through $\pi/4$ at the resonance 
point. 

Consider now the evolution of the $\nu_{eL}$'s in the sun. 
Since $\theta(r\simeq 0)\simeq 0$, it follows from eq. (\ref{eigen}) that 
a $\nu_{eL}$ born in the core of the sun nearly coincides with one of the 
neutrino matter eigenstates, namely $\nu_A$. If the matter density changes 
slowly enough (adiabatically) along the neutrino path, the neutrino system 
has enough time to adjust itself to the changing external conditions and the 
eigenstate $\nu_A$ will propagate in matter and magnetic field as $\nu_A$ 
without being converted into $\nu_B$. 
However, the composition of the eigenstates $\nu_A$ and $\nu_B$ changes 
along the neutrino path since the mixing angle $\theta$ changes. 
In particular, at the surface of 
the sun where $\theta\simeq \pi/2$ the $\nu_A$ practically coincides with 
the $\bar{\nu}_{\mu R}$. This means that we have a complete (or almost 
complete) adiabatic conversion of $\nu_{eL}$ into $\bar{\nu}_{\mu R}$ in the 
sun. This phenomenon is very similar to the $\nu_{eL}\to\nu_{\mu L}$ 
conversion due to the MSW effect. In fact, this is nothing but the well known 
Landau--Zener effect. 
The energy levels of a quantum mechanical 
system cross in the absence of the mixing interaction. However the mixing 
leads to an avoided level crossing. If the neutrino starts at high densities 
as the $\nu_{eL}$ which coincides with the higher--energy eigenstate $\nu_A$ 
and remains on the higher--energy branch while propagating to the low 
densities, it will end up as the $\bar{\nu}_{\mu R}$. 

Now, how can one quantify the adiabaticity condition? This is the condition 
that the external parameters change slowly enough along the neutrino 
trajectory so that the ``jumps'' between the $\nu_A$ and $\nu_B$ states 
are suppressed. One can show that the adiabaticity condition is most 
restrictive in the vicinity of the resonance. It can be formulated there as 
the requirement that at least one precession length $l_{prec}=\pi/\mu 
B_\bot$ fit into the resonance width $\Delta r$ which is defined as the 
spatial width of the resonance at half height. More precisely,
\be
\lambda_{\rm RSFP}\equiv \pi\frac{\Delta r}{l_{prec}}=8\frac{E}
{\Delta m^2}\,(\mu B_{\bot r})^2\,L_{\rm eff} > 1\,.
\label{adiab1}
\ee
Here $B_{\bot r}$ is the magnetic field strength at the resonace and 
$L_{\rm eff}=|1/N_{\rm eff}(dN_{\rm eff}/dr)|_r^{-1}$ is the effective matter 
density scale height (the distance over which $N_{\rm eff}$ varies 
significantly). In the sun, for $0.1 R_\odot \aprle r\aprle 0.9 R_\odot$, 
the scale height $L_{\rm eff}\simeq 0.1 R_\odot$. 
Notice that $\lambda_{\rm RSFP}$ is proportional to $(\mu B_\bot)^2$ so that 
the adiabaticity condition puts a lower bound on the product of the neutrino 
transition magnetic moment and magnetic field strength at the resonance. 

It is instructive to compare eq. (\ref{adiab1}) with the adiabaticity 
condition of the MSW effect:
\be
\lambda_{\rm MSW}=\frac{\sin^22\theta_0}{\cos 2\theta_0}\, 
\frac{\Delta m^2}{2E} L_{\rm eff} > 1\,.
\label{adiab2}
\ee
We see that the energy dependence of the two adiabaticity parameters is quite 
different: $\lambda_{\rm RSFP}\propto E$ whereas $\lambda_{\rm MSW}\propto 
E^{-1}$. Therefore one could hope to be able to experimentally distinguish 
the RSFP from the MSW effect by the distortions they cause to the solar 
neutrino spectrum. 
Unfortunately, in reality the situation is much more complicated. The 
RSFP adiabaticity parameter depends crucially on the magnetic field 
strength at the resonance $B_{\bot r}$. Since the solar magnetic field is 
not uniform, $B_\bot=B_\bot(r)$, the value $B_{\bot r}$ depends on the 
resonance coordinate which in turn depends on the neutrino energy $E$ 
through the resonance condition (\ref{res1}). Thus, the adiabaticity 
parameter $\lambda_{\rm RSFP}$ has an additional implicit neutrino energy 
dependence entering through $B_{\bot r}$. Since the profile of the 
solar magnetic field is essentially unknown the distortion of the 
energy spectrum of the solar neutrinos cannot be unambiguously predicted. 
In particular, one cannot be sure that the RSFP of neutrinos and the MSW 
effect would result in qualitatively different distortions of the neutrino 
spectrum. Still, we can learn something from this discussion.  
First, unlike the ordinary neutrino spin precession, the RSFP 
{\em does} depend on the neutrino energy. This in particular means that the 
RSFP should affect different solar neutrino experiments to a different extent 
since they are sensitive to different domains of the solar neutrino 
spectrum. Second, even though the neutrino spectrum distortion due to the 
RSFP is to a large extent uncertain, it is not absolutely arbitrary. The 
point is that not all the conceivable magnetic field profiles fit well the 
existing solar neutrino data. As I will discuss later, there is a limited 
number of the magnetic field profiles that can do the job and so the allowed 
neutrino spectrum distortions are also limited. 

\subsection{RSFP in twisting magnetic fields}
So far in my discussion 
I was assuming that the component of the magnetic field strength which is  
transverse to the neutrino momentum does not change its direction 
as the neutrino propagates. What happens if the magnetic 
field has different orientations in the transverse plane at different points 
$r$ along the neutrino path? Such ``twisting'' fields can lead to 
interesting consequences [11-17]. 
The mixing term in the evolution equation (\ref{evol1}) now depends on the 
angle $\phi(r)$ between the magnetic field and a fixed direction in the 
transverse plane: $V_{ij}=\mu B_\bot(r)\,e^{i\,\phi(r)}$. By moving into 
the reference frame which rotates with the magnetic field one can 
eliminate the $e^{i\,\phi(r)}$ factor in $V_{ij}$, but there will arise 
the additional $\pm \phi'(r)/2$
terms in the 
diagonal elements of the effective Hamiltonian in eq. (\ref{evol1}). Such 
terms can be conveniently considered as a modification of the effective 
matter density. The adiabaticity condition (\ref{adiab1}) will also be 
modified,  
it will depend on $\phi''(r)$. 
In general, a right-handed twist ($\phi'<0$) tends to enhance the RSFP 
whereas a left-handed twist tends to suppress it \cite{Sm,AKS}.     

Now the question is: When are the effects of the twisting fields 
of importance, or how large the twist $\phi'$ should be to affect the RSFP 
significantly? It is easy to see that the effects of the twisting fields 
become significant when 
\be
\phi'\sim \sqrt{2} G_F N_{\rm eff}\,. 
\label{twist}
\ee
Let us make a simple estimate. Assume $\phi' \sim 1/r_0$ where $r_0$ is 
the curvature radius of the magnetic field lines. Take for $N_{\rm eff}$ 
the value of the matter density at the bottom of the convective zone of the
sun. Then eq. (\ref{twist}) gives $r_0\sim 0.1 R_\odot$, which is quite a 
reasonable value. Thus, the effects of the field twist may be
significant for the solar neutrinos. I will come back to this point 
later when I discuss the Homestake data.

Twisting magnetic fields can lead to new interesting phenomena in neutrino 
physics, such as, e.g., new resonances \cite{Sm,AKS,APS}. 
Unfortunately, I don't have enough time to discuss this topic.

\subsection{Combined effects of neutrino oscillations and SFP}

Up to now I have been assuming that the usual neutrino mass mixing is 
absent and therefore neutrinos do not oscillate. However, the 
non-vanishing neutrino transition magnetic moments can exist only if the 
lepton flavor is not conserved. In such a situation it is natural to expect 
the mass mixing to be present as well. The above discussion is anyway valid 
if the vacuum neutrino mixing angle is not too large: $\tan 2\theta_0 <
\tan 2\theta (N_{\rm eff}=0)$. This condition is sufficient but, as we 
shall see, not always necessary. 
 
The combined action of the RSFP and the neutrino oscillations does not 
just reduce to a  mechanical 
sum of the two processes. 
It turns out that nontrivial new effects 
appear. Consider, for example, the 
energy level scheme for the neutrino system under the discussion (Fig. 1). 
\begin{figure}
\centerline{\psfig{figure=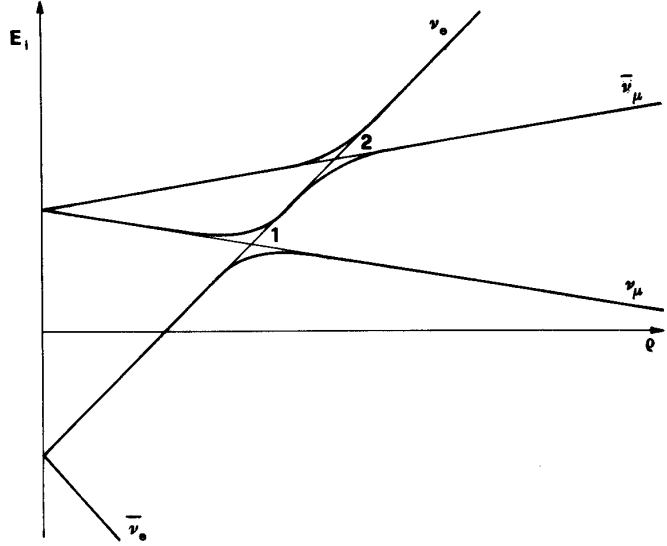,height=8.5cm}}
\caption{Neutrino energy levels vs matter density $\rho$ for the Majorana
($\nu_e, \nu_\mu$) system with mixing and transition magnetic moment. 1 -- MSW 
resonance, 2 -- RSFP resonance.}
\end{figure}
There are two avoided level crossings now, i.e. two resonances. The 
resonance of the SFP occurs at higher densities than the MSW resonance.
This can be easily understood if we compare the resonance conditions 
(\ref{res1}) and (\ref{res2}). 
Therefore a $\nu_{eL}$ born  
in the core of the sun and propagating outwards will first encounter the 
RSFP resonance and then the MSW one. However, if the adiabaticity 
condition of the RSFP is satisfied, the $\nu_{eL}$  will be adiabatically 
converted into the $\bar{\nu}_{\mu R}$ and so will never reach the MSW 
resonance. Thus, the MSW effect would be inoperative in this case, no matter 
how large the MSW adiabaticity parameter $\lambda_{\rm MSW}$. 

The above situation refers to the case when the two resonances are well 
separated. If they overlap, an interesting interplay of the MSW effect and 
the RSFP of neutrinos may occur. The two resonances may either enhance or 
suppress each other, depending on their degree of adiabaticity. A detailed 
discussion of this interplay can be found in 
[18-21]. 
Another important consequence of the joint operation of the neutrino 
oscillations and SFP is the possibility of the transformation of the solar 
$\nu_{eL}$'s into $\bar{\nu}_{eR}$'s \cite{LM,Akhm4,Akhm5,RBLBPP,BL1}.  
This may have very important experimental consequences which I will discuss 
later on. 
 
\section{Experimental data}
I will now briefly discuss the available experimental data. 
At the moment we have the data from the five solar neutrino experiments: 
Homestake, GALLEX, SAGE, Kamiokande and Super--Kamiokande. All the data show 
some deficiency of the solar neutrinos, the degree of which varies from the 
experiment to the experiment. This was reviewed in the other talks at this 
conference \cite{Exprev1,Exprev2,Exprev3}. 
I will concentrate on a different aspect of the data -- their time 
structure. I will not discuss the Super--Kamiokande data 
since it is too early yet to discuss its possible time dependence. 

As I have already mentioned, there are some indications of the time 
variations of the signal in the Homestake experiment. At the same 
time, the Kamiokande group did not observe any time variation of the solar 
neutrino signal in their experiment, which allowed them to put an upper limit 
on the possible time variation, $\Delta Q/Q < 30\%$ at 90\% c.l. 
\cite{Kamlimit}. The gallium experiments have not observed any statistically 
significant time variation of the signal either \cite{Exprev3}. 

Some remarks about time structure of the Homestake data are in order. 
The data compare better with an assumption of a time--dependent signal 
than with that of a constant one, hinting to an anticorrelation with solar 
activity. For the solar cycles 20 and 21 the detection rate at the 
periods of the quiet sun exceeded that for the active sun by a factor 
of two or more. There are also some indications in favor of the semiannual 
variations, but their statistical significance is lower than that for the 
11-year variations. There exist more than 20 analyses of the 
data which used a variety of statistical methods [29-53]. The authors of 
all these papers but three \cite{Lan,Wi,Liss} came to the conclusion that 
the Homestake data (or at least part of it) shows a time variation in 
anticorrelation with solar activity. However, the obtained values of the 
correlation coefficients and the confidence level depend very much on the 
indicator of solar activity chosen, the statistical methods used and on the 
way the experimental errors are treated. Moreover, almost all the analyses 
were performed before 1990 and so did not take into account more recent runs 
(run 109 and later). These runs  do not show a tendency to vary in time, 
similarly to the earlier runs 19--59. 
A recent analysis of Stanev \cite{St} which updated the one of ref. 
\cite{BSSS} included the runs 109-133 and showed that this results in the 
correlation coefficient between the Homestake detection rate and the 
sunspot number being significantly decreased. 

I would like to mention an interesting recent development in the 
analyses of the time structure of the Homestake data. Two groups 
\cite{Oak1,Obr,Oak2} pointed out that the total sunspot number is too 
rough an indicator of solar activity which only characterizes its gross 
features. They argued than one should use a more local indicator if one 
is interested in the correlation with the solar neutrino signals. They 
suggested to use the direct measurements of the magnetic field strength on 
the surface of the sun along the line of the sight connecting the center of 
the sun and the earth, i.e. along the trajectory of the solar 
neutrinos observed at the earth \cite{Oak1,Obr,Oak2}. The authors of these  
studies came to the conclusion that the (anti)correlation coefficient 
increases significantly if one uses this local characteristics of the 
solar activity. Another very interesting point was made by Massetti and 
Storini \cite{MaSt}. They found that the anticorrelation between the 
Homestake signal and the green corona line (which they chose as an estimator 
of the solar magnetic field) is much stronger for the neutrinos emitted 
in the southern solar hemisphere than for those emitted in the northern 
hemisphere. This conclusion was confirmed by Stanev \cite{St}. This result 
may have a very interesting interpretation. As was reported by Rust at 
this conference \cite{Rust}, about 80\% of the magnetic field ejected from  
the southern solar hemisphere has a right-handed twist ($\phi' <0$),  
while about 80\% of the field ejected from the northern hemisphere has a 
left-handed twist. As I mentioned before, 
the right-handed twist enhances the RSFP of neutrinos whereas the 
left-handed twist suppresses is. Therefore the observation of Massetti 
and Storini finds a very natural interpretation in the framework of the
neutrino transition magnetic moment scenario (see \cite{Sm,AKS} for an early 
discussion of possible north-south effects due to the twisting magnetic 
fields). There is one more observation that supports this interpretation. As 
was noticed by Stanev \cite{St}, two solar cycles ago there was no significant 
time variation in the Homestake data (runs 19--59), similar to what was 
observed in 1990--1996. This may indicate that the Homestake data has a 
better 22-year periodicity than the 11-year one. Since the magnitude of the 
solar magnetic field changes with the 11-year periodicity whereas its 
$direction$ (including the twist) changes with the 22-year periodicity, this 
might be an additional argument for the RSFP in twisting magnetic fields as 
the solution of the solar neutrino problem.

The question of whether or not any time variations have been observed by the 
Homestake experiment is rather controversial. Unfortunately, the statistics 
of the Homestake experiment is poor and it is very difficult (if 
possible at all) to draw any definitive conclusion regarding the time 
dependence of the data. However, in any case the following question 
naturally arises: Can one reconcile, at least in principle, a strong time 
variation in the Homestake data with no observable time variation in 
Kamiokande and GALLEX experiments? I will show now that the answer to 
this question is yes, at least within the RSFP scenario. The RSFP mechanism 
can very naturally account for all the available solar neutrino data. Let me 
now briefly describe how this works. 

\section{Reconciling the time structure of the data} 
Suppose that we believe in strong (by a factor of two or more) time 
variations in the Homestake data. If we are to describe the absence of the
observable time variations in the Kamiokande and gallium experiments, the 
mechanism responsible for the time variations 
must be neutrino energy dependent, otherwise the degree of the time   
variations in all these experiments would be the same. This excludes the 
ordinary neutrino spin precession which in energy independent. By contrast, 
the RSFP mechanism $is$ energy dependent, so it has at least a potential of 
accounting for the data. We shall now see that how this potential is 
realized.
\begin{figure}
\centerline{\psfig{figure=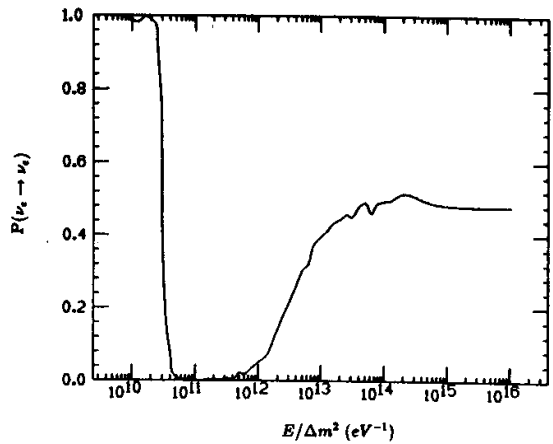,height=6.7cm}}
\caption{Typical RSFP ``suppression pit''. This figure was borrowed
from [58]. }
\end{figure}

In many respects the RSFP of neutrinos is similar to the MSW effect. 
In particular, the $\nu_{eL}$ survival probability $P$ as the 
function of $E/\Delta m^2$ which determines 
the energy dependence of the effect is given by a ``suppression pit''. 
A typical RSFP suppression pit is shown in Fig. 2. Its shape is similar to 
the MSW suppression pit, although there are some differences. For example, in 
the case of the MSW effect the r.h.s edge of the curve approaches unity, 
whereas for the RSFP it approaches 1/2. It has been emphasized in \cite{LN} 
that the RSFP--type shape of the suppression pit can explain very 
naturally the observed average depletions of the solar neutrino signal 
observed  in different experiments (even if one forgets about any time 
variations): the low-energy $pp$ neutrinos are unsuppressed ($P
\approx 1$ at the l.h.s. edge of the curve), the intermediate energy 
$^7$Be and $^8$B neutrinos are at the bottom of the suppression pit and 
so are strongly suppressed, whereas the high-energy part of the $^8$B 
neutrino spectrum is on the r.h.s. edge of the pit and therefore is 
suppressed by a factor of two. This is exactly what one needs to account 
for the experimental data. Notice that the same can be achieved in the 
case of the MSW effect by a rather fine adjustment of the $\Delta m^2$ 
parameter in order to place the high energy part of the $^8$B neutrino 
spectrum on the r.h.s. slope of the MSW suppression pit in such a way as 
to achieve a factor of two suppression on the average. In the case of the 
RSFP mechanism this comes out quite naturally and for a wider range of the 
values of the $\Delta m^2$ parameter. 

\subsection{GALLEX and SAGE}
Let us see now discuss how the RSFP mechanism explains the lack of the 
time dependence of the signal in the gallium experiments. 

The shape of the suppression pit depends on the profile and strength of the 
magnetic field of the sun. In particular, the depth of the pit depends on 
the maximum field strength while its width depends on the extension of solar 
magnetic field. Low energy neutrinos experience the RSFP conversion  
at high densities since the resonance density is inversely proportional 
to the neutrino energy [see eq. (\ref{res1})]. Therefore they are 
sensitive to the magnetic field strength in the core of the sun. If such 
an inner magnetic field is absent or weak, the RSFP will not be 
effective for the low energy neutrinos and their survival probability 
will be close to unity. For this reason the position of the l.h.s. edge of 
the suppression pit in Fig. 2 (low--energy edge) depends on the strength of 
the inner magnetic field of the sun. We do not know if a strong magnetic 
field exists in the core of the sun; however, it is known that in any 
case, if exists, it must be practically constant in time. 

Before the data of GALLEX and SAGE became available, one could had asked 
the following question: If one believes in the time variations in the 
Homestake data, can one predict the time structure of the signal in the 
gallium solar neutrino experiments? Riccardo Barbieri repeatedly asked me 
this question when I visited Pisa in 1989. The answer was given in 
\cite{Akhm6}. This answer was conditional, depending on the average 
suppression of the signal, which, of course, was unknown at that time:
If the average suppression of the signal is strong, there should be 
significant time variations of the signal, while if the average 
suppression is not too strong, no appreciable time variations should be 
detected. The reason for that is very simple. The inner magnetic field 
is unknown but in any case does not vary in time. If it is very strong, 
it may result in a significant (constant in time) suppression of the 
low-energy $pp$ neutrinos. In this case the time variations of the 
$^8$B neutrino contribution would be visible. On the other hand, if the 
inner magnetic field is weak, the $pp$ neutrino flux will leave the sun  
unscathed and constitute the major part of the signal. In this case   
even a strong time variation of the $^8$B neutrino flux will hardly be 
detectable since these neutrinos would constitute only a small fraction 
of the signal. We see now that this prediction is fully consistent with the
results of GALLEX and SAGE. They imply that the $pp$ neutrino flux is 
unsuppressed; therefore within the RSFP scenario one should not see 
any appreciable time variations. As a matter of fact, no statistically 
significant time variations of the signal were observed. Thus, the results 
of the GALLEX and SAGE experiments can be easily fitted within the RSFP 
mechanism provided that the magnetic field in the core of the sun $B_i$ 
is not too strong. One can turn the argument around and ask the following 
question: If we believe in the RSFP mechanism, what is the maximum allowed 
inner magnetic field which is not in conflict with the gallium experiments? 
The answer turns out to be $(B_i)_{max}\approx 3\times 10^6$ G assuming the 
neutrino transition magnetic moment $\mu=10^{-11}\mu_B$ \cite{ALP1}. 

\subsection{Kamiokande}
The question I am going to address now is how to reconcile strong 
time dependence of the signal in the Homestake experiment with no or  
little time variation of the Kamiokande data. I will show that this comes 
out quite naturally in the RSFP scenario. The key points here are that 
[56-59] 

(1) The two experiments are sensitive to slightly different parts of the 
solar neutrino spectrum: Homestake is sensitive to the energetic ${}^8$B 
neutrinos as well as to medium--energy ${}^7$Be and $pep$ neutrinos, 
whereas the Kamiokande experiment is only sensitive to the high--energy 
part of ${}^8$B neutrinos ($E>7.5$ MeV); 

(2) For Majorana neutrinos, the RSFP converts left-handed $\nu_{e}$ into 
right-handed $\bar{\nu}_{\mu}$ (or $\bar{\nu}_{\tau}$) which are sterile for 
the Homestake experiment (since their energy is less than the muon or tauon 
mass) but do contribute to the event rate in the Kamiokande experiment 
through their neutral--current interaction with electrons. Although the 
$\bar{\nu}_{\mu}e$ cross section is smaller than the $\nu_{e}e$ one, it is 
non-negligible, which reduces the amplitude of the time variation of the  
signal in the Kamiokande experiment. It turns out that the above two points 
are enough to account for the differences in the time dependences of the 
signals in the Homestake and Kamiokande experiments. 
\begin{figure}
\centerline{\psfig{figure=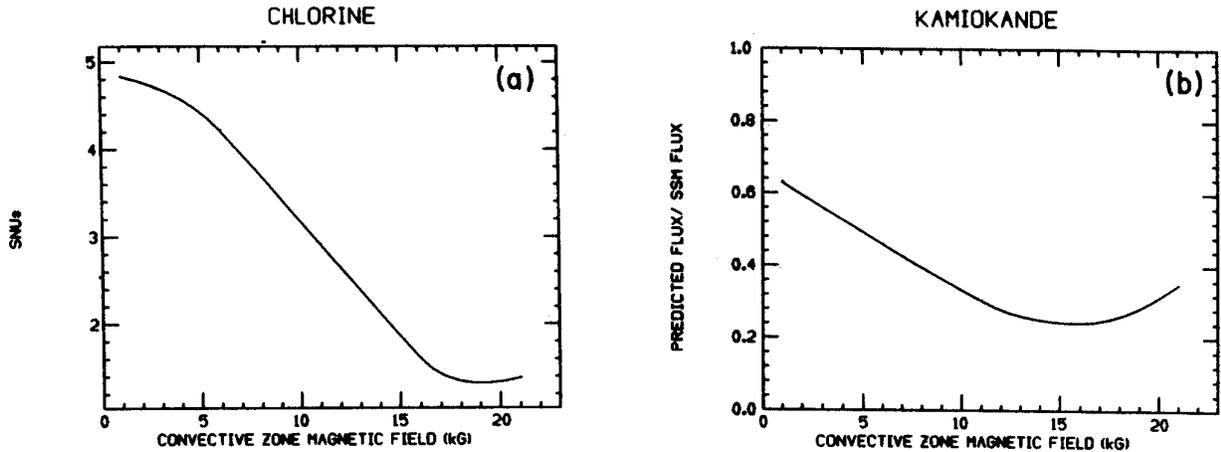,height=6.7cm}}
\caption{The calculated Homestake and Kamiokande detection rates vs
convective zone magnetic field strength [58]. The neutrino transition
magnetic moment $\mu=2\times 10^{-11} \mu_B$ is assumed.}
\end{figure}

In Fig. 3 are shown the event rates in the Homestake 
and Kamiokande experiments as functions of the convective zone magnetic 
field strength calculated by Babu, Mohapatra and Rothstein \cite{BMR}. 
There are three important observations to be made. First, both rates 
decrease with magnetic field until they reach their minima, and then start 
to increase. Second, the minimum of the Kamiokande signal is situated at a 
lower magnetic field strength then the one of the Homestake signal.  Third, 
the Kamiokande curve is less steep than the Homestake one. Let me explain 
this behavior of the curves. The third of the above observations 
obviously follows from the above point (2). For weak fields, the 
efficiency of the RSFP increases with the field (so that the counting rates 
decrease) because the adiabaticity of the transition improves. The situation 
changes, however, after the field becomes strong enough. As I discussed  
above, a $\nu_{eL}$ can be fully converted into the $\bar{\nu}_{\mu R}$ if 
it is born at a density much above the resonance one and passes through the 
RSFP resonance adiabatically. However, if it is born close to the resonance, 
the efficiency of the conversion decreases; for example, for the $\nu_{eL}$ 
born exactly at the resonance point, the conversion probability is only 1/2 
even in the case of a perfectly adiabatic conversion. The resonance width 
$\Delta r$ is 
\be
\Delta r = \frac{2\mu B_{\bot r}}{\Delta m^2/2E}\,L_{\rm eff}\,,
\label{deltar}
\ee
so it increases with the field strength. For very strong magnetic fields 
the resonance becomes very wide, so the neutrino birth place becomes too 
close to the resonance region and the efficiency of the conversion 
decreases. This explains the existence of the minima of the detection 
rates in Fig. 3. From the above arguments it follows that there is an 
optimum value (or a range of values) of the resonance widths $\Delta r
\simeq const$ which corresponds to the maximum efficiency of the RSFP, 
i.e. to the minima of the detection rates. Since $\Delta r \propto 
B_\bot E$, the higher the neutrino energy, the smaller the magnetic 
field strength at the minimum of the detection rate. This explains why 
the detection rate for the the Kamiokande experiment (which is sensitive    
to more energetic neutrinos) has its minimum situated at a lower magnetic 
field strength than that for the Homestake experiment. 

Now let us look again at Fig. 3. Assume that during the solar cycle the 
convective field magnetic field changes, say, between 10 and 20 kG. 
For this range of $B_c$ the Homestake detection rate changes steeply. 
At the same time, the Kamiokande rate is close to 
its minimum and so varies very little. It is for this reason that  
the time variation of the Kamiokande signal may be much smaller than that 
of the Homestake signal. One can draw another important conclusion from the 
above discussion: {\em The time structure of the Kamiokande (and now 
Super--Kamiokande) experiment may be more complicated than a mere 
anticorrelation with the solar activity. In particular, one can expect 
an anticorrelation during a part of the solar cycle and direct  
correlation during another part}. This can be tested in the 
Super--Kamiokande experiment.  

\section{Fitting the data}
I will now discuss more quantitatively how one fits the data. 
The main problem one encounters is that the magnetic field inside the 
sun is practically unknown. In the case of the MSW effect, all the 
experimental data are analyzed in terms of the two unknown parameters: 
$\Delta m^2$ and $\sin^2 2\theta_0$. In the case of the RSFP 
mechanism the situation is much more complicated. One has an unknown 
parameter $\Delta m^2$ and an unknown function -- magnetic field profile 
in the sun $B_{\bot}(r)$ (in fact, there is one more parameter -- the 
neutrino transition magnetic moment, but it enters the problem only  
through the product $\mu B_\bot$). In such a situation the only thing one 
can do is to try various ``plausible'' magnetic field configurations, 
calculate the detection rates for various solar neutrino experiments 
and compare them with the experimental results. Such a program was 
carried out in a number of papers (see, e.g., [15-22, 55, 57-64]). 
I will discuss 
briefly the results of the two recent papers \cite{ALP1,ALP2}.
In \cite{ALP1} nine different model magnetic field profiles were used. 
The neutrino mass mixing was assumed to be absent. Two of the nine magnetic 
field profiles used reproduced the data very well, while the others either 
produced poor quality fits or completely failed to reproduce the data. This 
shows that one can in principle discriminate between various models of the 
solar magnetic field using the solar neutrino data. The upper limit on the 
inner magnetic field which I already mentioned before was established. 
The typical values of the parameters that produced the good fits of the data 
were $\Delta m^2\simeq (4\times 10^{-9}-2\times 10^{-8})$ eV$^2$, with the 
maximum magnetic field strength varying between $\simeq 25$ and 
$\simeq 50$ kG during the solar cycle (assuming $\mu=10^{-11}\mu_B$).   

In \cite{ALP2} it was assumed that the neutrino mass mixing is present 
($\theta_0\ne 0$) i.e. the combined effect of the RSFP and neutrino 
oscillations was studied. 
The neutrino signals in the chlorine, gallium and Kamiokande experiments 
were calculated using nine different model magnetic field profiles.  
Three of the nine magnetic field configurations used produced good fits of 
all the data. Typical values of the neutrino parameters required to 
account for the data are $\Delta m^2 \simeq (10^{-8}$--$10^{-7})$ eV$^2$, 
$\sin 2\theta_0 \aprle$ 0.25; larger mixing angles are excluded by the 
GALLEX and SAGE data. 
For neutrino transition magnetic moment $\mu=10^{-11}\mu_B$ 
the maximum magnetic field in the solar convective zone has to vary in time 
in the range (15--30) kG. 

As I have already mentioned, 
if neutrinos experience the RSFP in the sun and also have 
mass mixing, a flux of electron antineutrinos can be produced. 
This flux is in principle detectable in the SNO, Super--Kamiokande and 
Borexino experiments even in the case of moderate neutrino mixing angles  
\cite{LM,Akhm4,Akhm5,RBLBPP,BL1}. 
The main mechanism of the $\bar{\nu}_{eR}$ production is $\nu_{eL}\rightarrow 
\bar{\nu}_{\mu R}\rightarrow \bar{\nu}_{eR}$, where the first transition 
is due to the RSFP in the sun and the second one is due to the vacuum 
oscillations of antineutrinos on their way between the sun and the earth. 
The salient feature of this flux is that it should vary in time in 
anticorrelation with the variation of the $\nu_{eL}$ flux. The detection of 
the solar $\bar{\nu}_{eR}$ flux would be a signature of the combined effect 
of the RSFP and neutrino oscillations. It could allow one to discriminate 
between small mixing angle ($\sin 2\theta_0\aprle 0.1$) and moderate mixing 
angle ($\sin 2\theta_0\simeq 0.1 - 0.25$) solutions. For the 
mixing angles $\sin 2\theta_0\aprge 0.3$, a large flux of electron 
antineutrinos would be produced in contradiction with an upper limit 
already derived from the Kamiokande and LSD data \cite{BFMM,LSD}. 

The $\bar{\nu}_{eR}$ flux can be significantly 
enhanced if the solar magnetic field changes its direction along the 
neutrino trajectory, i.e. twists \cite{APS,BL2}. In this case one can 
have a detectable $\bar{\nu}_{eR}$ flux even if the neutrino magnetic moment 
is too small or the solar magnetic field is too weak to account for the solar 
neutrino problem \cite{BL2}. 

\section{Predictions for future experiments}
The predictions of the RSFP mechanism for the Super--Kamiokande experiment 
are very similar to the ones for Kamiokande: as I discussed above, one can 
expect small time variations, not necessarily in anticorrelation with the 
solar activity. In fact, during a part of the solar cycle the signal may 
vary in direct correlation with the solar magnetic field. Similar prediction 
is made for the charged--current (CC) signal in SNO. However, in that case 
one should expect somewhat stronger time variations since the  
$\bar{\nu}_{\mu R}$ or $\bar{\nu}_{\tau R}$ produced as a result of the 
RSFP do not contribute to the CC signal. The neutral--current signal in 
SNO should remain constant in time (unless neutrinos precess into sterile 
states, i.e. are Dirac particles) 
modulo the trivial 7\% seasonal 
variations of the signal because of the varying distance between 
the sun and the earth. This along with the varying CC signal would be a 
clear signature of the neutrino spin-flavor precession.   

The $^7$Be neutrino flux is expected to be strongly suppressed; it should 
also vary in time, which may in principle be observable in the Borexino 
experiment. 
The expected variation of the $^7$Be neutrino flux is about 30\% 
(40\%) for $\sin 2\theta_0\simeq 0.24$ (0) \cite{ALP2}. 

For not too small neutrino mass mixing $(\sin 2\theta_0\aprge 0.1)$, one 
can also expect a flux of the $\bar{\nu}_{eR}$ coming from the sun, with 
a distinct feature of time variation in an anti-phase with the solar 
$\nu_{eL}$ flux. Such a flux may be observable in the SNO, Super--Kamiokande 
and Borexino experiments \cite{ALP2}.  

Even though the predicted time variation for Super--Kamiokande (and to some 
extent also for SNO) are smaller than the variations in the Homestake 
experiments, its amplitude cannot be smaller than 10--15\% provided that one 
believes in the time variations of the Homestake signal by a factor of 
two or more. Such a variation should be clearly detectable in the 
Super--Kamiokande. The amplitude of time variation of the charged-current 
signal in SNO should be even larger. 

Thus, the RSFP mechanism leads to very distinct and testable predictions.  

\section{{\em Pro et Contra}}
I will summarize now the arguments for and against the RSFP mechanism as 
the solution of the solar neutrino problem and comment on them. I start with 
the objections to this mechanism that are frequently put forward. 

\noindent
$\bullet$ {\em There isn't any time variation in the Homestake data; 
therefore there is nothing to discuss}. 

I will not comment on this point. 

\noindent
$\bullet$ {\em There is no compelling evidence for time variation of the 
Homestake signal; the errors are too large to draw any definitive 
conclusions.} 

This is certainly correct. Fortunately, the Super--Kamiokande has (and SNO 
is expected to have) a much higher detection rate than all the previous 
experiments; therefore one can hope that the question of whether or not 
the observed solar neutrino signal varies in time will be settled in a 
few years.  

\noindent
$\bullet$ {\em Kamiokande has not observed any time variation of the
signal during more than eight years of data taking; this disproves 
transition magnetic moment scenario.} 

This conclusion is just wrong, as I explained in detail above. 

\noindent
$\bullet$ {\em GALLEX and SAGE have not observed any time variation 
of the signal. Moreover, these experiments did not detect strong 
suppression of the signal during the period of high solar activity, which 
contradicts the transition magnetic moment scenario.} 

This is also wrong and was also discussed above. 

\noindent
$\bullet$ {\em Neutrino magnetic moments are severely constrained from 
above by astrophysics data; solar magnetic field is not strong enough to 
produce any sizeable spin or spin--flavor precession.} 

For the maximum magnetic field of the sun of the order of a few tens kG 
the requisite value of the neutrino transition magnetic moment $\mu\simeq 
10^{-11}\mu_B$. This is consistent with the existing laboratory upper bounds 
on $\mu$ and also with the astrophysics bounds coming from the white dwarf 
cooling rates. There are also a factor of 3 to 10 more stringent bounds 
coming from the cooling rates of the helium stars. However, at this level 
the bounds become less reliable. In any case, a factor of ten smaller $\mu$ 
would require a factor of ten stronger magnetic fields for the RSFP to 
proceed with the same efficiency. There are even more stringent bounds 
on neutrino magnetic moments coming from the neutrino signal observations 
from the supernova 1987A. However, these bounds are not applicable to the 
transition magnetic moments of Majorana neutrinos. 
The reason is very simple: These bounds are based on the assumption that 
neutrinos can be converted into right-handed sterile states in the 
supernova because of scattering on the particles of the matter through 
the photon exchange. Sterile neutrinos then escape freely from the dense core 
of the supernova instead of being trapped. However, in the case of 
Majorana neutrinos the resulting right-handed states are right-handed 
antineutrinos which are not sterile and therefore are still trapped.

The second part of the above objection, as well as next two, concerns the 
solar magnetic field. I will comment on them altogether. 

\noindent
$\bullet$ {\em Solar magnetic field is only strong in sunspots which occupy 
a small part of the solar surface; there is no strong enough large-scale 
magnetic field in the sun.} 

\noindent
$\bullet$ {\em The equatorial gap in the toroidal magnetic field of the sun
is rather wide, so that the year-averaged field affecting the solar neutrinos
should be small}. 

Unfortunately, the magnetic field inside the sun is not accessible to 
direct observations and is very poorly known. At the moment, there is no 
compelling model of the solar magnetic field (in particular, there is no 
model capable of {\em predicting} the 11-year period of the solar cycle).
Probably, this is related to the fact that the magnetic field plays a 
relatively minor role in the solar dynamics. The sole robust upper limit 
on the strength of the solar magnetic field comes from the requirement 
that the field pressure be smaller than the matter pressure; this is a 
rather weak limit (for the convective zone, $B\aprle 10^7$ G). There are 
more stringent bounds, 
but they are highly model dependent. Regarding the equatorial gap in the 
toroidal magnetic field of the sun. It is not very well defined; there are 
even periods when groups of the sunspots cross the solar equator. In any 
case, even if such a gap exist on the solar surface, it is not clear if it 
exists, e.g., near the bottom of the convective zone. To summarize, I think 
it is fair to say that at the moment there is a better chance to learn 
something about the solar magnetic field by studying the solar neutrinos 
than vice versa. 

\noindent
$\bullet$ {\em It is difficult to construct a particle-physics model which
would naturally produce large enough magnetic moments while keeping neutrino
mass small.} 

This is correct, such models are not easy to construct. In the standard 
model of electroweak interactions the magnetic moment of electron neutrino 
is $\mu_{\nu_e}\approx 3\times 10^{-19} \mu_B\,(m_{\nu_e}/{\rm eV})$, 
far too small to be of relevance for the solar 
neutrino problem. It is not difficult to construct an extension of the 
standard model with large enough $\mu_{\nu_e}$ or transition moment 
$\mu$; what is really difficult is to keep the neutrino mass naturally 
small in such a model. By ``naturally'' I mean to have neutrino mass small 
without introducing additional fine tuning of the parameters in the model. 

The standard electroweak model is full of fine-tuning, or hierarchy, 
problems (for example, why is $m_e$ more than 5 orders of magnitude smaller 
than the top quark mass?), so one could take an attitude that all these 
hierarchy problems will one day be solved altogether by a more general 
model which will encompass the standard electroweak model.  
However, it would be good to avoid introducing new fine tuning problems in 
the model if it is possible. Interestingly, there exist 
several successful attempts to construct such models, some of them very 
appealing (see, e.g., \cite{BM,BFZ}). Most of these models use various 
versions of the so-called Voloshin's symmetry \cite{Vol} to keep neutrino 
mass small without invoking a fine tuning. 

Now about the arguments supporting the RSFP scenario. I will mention only 
one of them: {\em The RSFP mechanism is fully consistent with all the 
currently available solar neutrino data}. 

Let me conclude with two remarks.
Both neutrino oscillations and spin (or spin--flavor) precession require 
some extension of the standard model of the electroweak interactions: in 
the {\em minimal} standard model neutrinos are massless and have zero 
magnetic moments. The most popular models of the neutrino mass invoke the 
so-called see-saw mechanism which requires some new physics at a very high 
energy scale, typically at the GUT scale $\sim 10^{16}$ GeV or 
``intermediate'' scale $\sim 10^{12}$ GeV. Obviously, this new physics cannot 
be directly probed.  On the other hand, all the extensions of the standard 
electroweak model which are capable of producing 
$\mu_{\nu}\sim (10^{-12}-10^{-11})\mu_B$ require some new physics at the 
energy scale $\sim$ a few hundreds GeV, i.e. close to the electroweak scale. 
This physics will hopefully become accessible to direct experimental tests in 
the near future.

In the beginning of the solar neutrino studies, when the first experiment 
was discussed and planned, the main idea was to probe the sun using the 
neutrinos it emits as a tool. Neutrinos themselves were considered 
standard and very well understood. Now the situation is quite different. 
It seems that now we know much more about the sun than we know about the 
neutrinos, so we can use the sun as a unique laboratory to study the 
neutrinos. In doing so, we should explore the neutrino 
properties in a broadest possible way, 
trying to get as much as possible information about all the neutrino 
characteristics that can be probed.
The neutrino magnetic moment is one of them. 

\section*{Acknowledgements}

I am grateful to the organizers of the 4th International Solar Neutrino 
Conference in Heidelberg for organizing a very interesting and enjoyable 
conference. Thanks are also due to John Bahcall for urging me to write a 
summary of the present status of the RSFP scenario.

\end{document}